\documentclass[twocolumn]{aastex631}
\usepackage{booktabs}
\usepackage{filecontents,catchfile}
\usepackage{longtable}
\usepackage{cleveref}
\usepackage{CJKutf8}

\newcommand{\RN}[1]{%
  \textup{\uppercase\expandafter{\romannumeral#1}}%
}

\newcommand{\um}{\,\micron}
\newcommand{\NeII}{[Ne\,\textsc{ii}]}
\newcommand{\NeIII}{[Ne\,\textsc{iii}]}
\newcommand{\SIV}{[S\,\textsc{iv}]}

\newcommand{\HII}{H\,\textsc{ii}}

\newcommand{\Lpahthree}{L\ensuremath{_\mathrm{PAH\,3.3}}}

\newcommand{\Ltotpah}{L\ensuremath{_\mathrm{\Sigma\,PAH}}}

\newcommand{\JWST}{\textit{JWST}}

\renewcommand{\thefootnote}{\fnsymbol{footnote}}

\newcommand{\paperI}{Paper~\textsc{i}}

\graphicspath{{./}{figures/}}

\shorttitle{Small neutral grains in NGC 7469}
\shortauthors{Lai et al.}

\begin{document}
\begin{CJK*}{UTF8}{bsmi}
\title{GOALS-JWST: Small neutral grains and enhanced 3.3 micron PAH emission in the Seyfert galaxy NGC 7469}

\correspondingauthor{Thomas S.-Y. Lai}
\email{ThomasLai.astro@gmail.com}

\author[0000-0001-8490-6632]{Thomas S.-Y. Lai (賴劭愉)}
\affil{IPAC, California Institute of Technology, 1200 E. California Blvd., Pasadena, CA 91125}

\author[0000-0003-3498-2973]{Lee Armus}
\affil{IPAC, California Institute of Technology, 1200 E. California Blvd., Pasadena, CA 91125}

\author[0000-0002-6570-9446]{Marina Bianchin}
\affiliation{Universidade Federal de Santa Maria, Departamento de Física, Centro de Ciências Naturais e Exatas, 97105-900, Santa Maria, RS, Brazil}
\affiliation{Department of Physics and Astronomy, 4129 Frederick Reines Hall, University of California, Irvine, CA 92697, USA}

\author[0000-0003-0699-6083]{Tanio D\'iaz-Santos}
\affiliation{Institute of Astrophysics, Foundation for Research and Technology-Hellas (FORTH), Heraklion, 70013, Greece}
\affiliation{School of Sciences, European University Cyprus, Diogenes street, Engomi, 1516 Nicosia, Cyprus}

\author[0000-0002-1000-6081]{Sean T. Linden}
\affiliation{Department of Astronomy, University of Massachusetts at Amherst, Amherst, MA 01003, USA}

\author[0000-0003-3474-1125]{George C. Privon}
\affiliation{National Radio Astronomy Observatory, 520 Edgemont Rd, Charlottesville, VA, 22903, USA}
\affiliation{Department of Astronomy, University of Virginia, 530 McCormick Road, Charlottesville, VA 22903, USA}
\affiliation{Department of Astronomy, University of Florida, P.O. Box 112055, Gainesville, FL 32611, USA}

\author[0000-0003-4268-0393]{Hanae Inami}
\affiliation{Hiroshima Astrophysical Science Center, Hiroshima University, 1-3-1 Kagamiyama, Higashi-Hiroshima, Hiroshima 739-8526, Japan}

\author[0000-0002-1912-0024]{Vivian U}
\affiliation{Department of Physics and Astronomy, 4129 Frederick Reines Hall, University of California, Irvine, CA 92697, USA}

\author[0000-0002-4375-254X]{Thomas Bohn}
\affiliation{Hiroshima Astrophysical Science Center, Hiroshima University, 1-3-1 Kagamiyama, Higashi-Hiroshima, Hiroshima 739-8526, Japan}

\author[0000-0003-2638-1334]{Aaron S. Evans}
\affiliation{National Radio Astronomy Observatory, 520 Edgemont Rd, Charlottesville, VA, 22903, USA}
\affiliation{Department of Astronomy, University of Virginia, 530 McCormick Road, Charlottesville, VA 22903, USA}

\author[0000-0003-3917-6460]{Kirsten L. Larson}
\affiliation{AURA for the European Space Agency (ESA), Space Telescope Science Institute, 3700 San Martin Drive, Baltimore, MD 21218, USA}

\author[0000-0001-7449-4638]{Brandon S. Hensley}
\affiliation{Department of Astrophysical Sciences, Princeton University, Princeton, NJ 08544, USA}

\author[0000-0003-1545-5078]{J.-D.T. Smith}
\affiliation{Ritter Astrophysical Research Center, University of Toledo, Toledo, OH 43606, USA}

\author[0000-0001-6919-1237]{Matthew A. Malkan}
\affiliation{Department of Physics \& Astronomy, 430 Portola Plaza, University of California, Los Angeles, CA 90095, USA}

\author[0000-0002-3139-3041]{Yiqing Song}
\affiliation{European Southern Observatory, Alonso de Córdova, 3107, Vitacura, Santiago, 763-0355, Chile}
\affiliation{Joint ALMA Observatory, Alonso de Córdova, 3107, Vitacura, Santiago, 763-0355, Chile}

\author[0000-0002-2596-8531]{Sabrina Stierwalt}
\affiliation{Physics Department, 1600 Campus Road, Occidental College, Los Angeles, CA 90041, USA}

\author[0000-0001-5434-5942]{Paul P. van der Werf}
\affiliation{Leiden Observatory, Leiden University, PO Box 9513, 2300 RA Leiden, The Netherlands}

\author[0000-0002-6149-8178]{Jed McKinney} 
\affiliation{Department of Astronomy, University of Massachusetts, Amherst, MA 01003, USA.}

\author[0000-0002-5828-7660]{Susanne Aalto}
\affiliation{Department of Space, Earth and Environment, Chalmers University of Technology, 412 96 Gothenburg, Sweden}

\author{Victorine A. Buiten}
\affiliation{Leiden Observatory, Leiden University, PO Box 9513, 2300 RA Leiden, The Netherlands}

\author[0000-0002-5807-5078]{Jeff Rich}
\affiliation{The Observatories of the Carnegie Institution for Science, 813 Santa Barbara Street, Pasadena, CA 91101}

\author[0000-0002-2688-1956]{Vassilis Charmandaris}
\affiliation{Department of Physics, University of Crete, Heraklion, 71003, Greece}
\affiliation{Institute of Astrophysics, Foundation for Research and Technology-Hellas (FORTH), Heraklion, 70013, Greece}
\affiliation{School of Sciences, European University Cyprus, Diogenes street, Engomi, 1516 Nicosia, Cyprus}

\author[0000-0002-7607-8766]{Philip Appleton}
\affiliation{IPAC, California Institute of Technology, 1200 E. California Blvd., Pasadena, CA 91125}

\author[0000-0003-0057-8892]{Loreto Barcos-Mu\~noz}
\affiliation{National Radio Astronomy Observatory, 520 Edgemont Rd, Charlottesville, VA, 22903, USA}
\affiliation{Department of Astronomy, University of Virginia, 530 McCormick Road, Charlottesville, VA 22903, USA}

\author[0000-0002-5666-7782]{Torsten B\"oker}
\affiliation{European Space Agency, Space Telescope Science Institute, Baltimore, MD 21218, USA}

\author[0000-0002-1392-0768]{Luke Finnerty}
\affiliation{Department of Physics \& Astronomy, 430 Portola Plaza, University of California, Los Angeles, CA 90095, USA}

\author[0000-0002-6650-3757]{Justin A. Kader}
\affiliation{Department of Physics and Astronomy, 4129 Frederick Reines Hall, University of California, Irvine, CA 92697, USA}

\author[0000-0002-9402-186X]{David R.~Law}
\affiliation{Space Telescope Science Institute, 3700 San Martin Drive, Baltimore, MD 21218, USA}

\author[0000-0001-7421-2944]{Anne M. Medling}
\affiliation{Department of Physics \& Astronomy and Ritter Astrophysical Research Center, University of Toledo, Toledo, OH 43606,USA}

\author[0000-0002-1207-9137]{Michael J. I. Brown}
\affiliation{School of Physics and Astronomy, Monash University, Clayton, VIC 3800, Australia}

\author[0000-0003-4073-3236]{Christopher C. Hayward}
\affiliation{Center for Computational Astrophysics, Flatiron Institute, 162 Fifth Avenue, New York, NY 10010, USA}

\author[0000-0001-6028-8059]{Justin Howell}
\affiliation{IPAC, California Institute of Technology, 1200 E. California Blvd., Pasadena, CA 91125}

\author[0000-0002-4923-3281]{Kazushi Iwasawa}
\affiliation{Institut de Ci\`encies del Cosmos (ICCUB), Universitat de Barcelona (IEEC-UB), Mart\'i i Franqu\`es, 1, 08028 Barcelona, Spain}
\affiliation{ICREA, Pg. Llu\'is Companys 23, 08010 Barcelona, Spain}

\author[0000-0003-2743-8240]{Francisca Kemper}
\affiliation{Institut de Ciencies de l'Espai (ICE, CSIC), Can Magrans, s/n, 08193 Bellaterra, Barcelona, Spain}
\affiliation{ICREA, Pg. Lluís Companys 23, Barcelona, Spain}
\affiliation{Institut d'Estudis Espacials de Catalunya (IEEC), E-08034 Barcelona, Spain}

\author[0000-0001-7712-8465]{Jason Marshall}
\affiliation{4Glendale Community College, 1500 N. Verdugo Rd., Glendale, CA 91208}

\author[0000-0002-8204-8619]{Joseph M. Mazzarella}
\affiliation{IPAC, California Institute of Technology, 1200 E. California Blvd., Pasadena, CA 91125}

\author[0000-0002-2713-0628]{Francisco Müller-Sánchez}
\affiliation{Department of Physics and Materials Science, The University of Memphis, 3720 Alumni Avenue, Memphis, TN 38152, USA}

\author[0000-0001-7089-7325]{Eric J. Murphy}
\affiliation{National Radio Astronomy Observatory, 520 Edgemont Rd, Charlottesville, VA, 22903, USA}

\author[0000-0002-1233-9998]{David Sanders}
\affiliation{Institute for Astronomy, University of Hawaii, 2680 Woodlawn Drive, Honolulu, HI 96822}

\author[0000-0001-7291-0087]{Jason Surace}
\affiliation{IPAC, California Institute of Technology, 1200 E. California Blvd., Pasadena, CA 91125}



\begin{abstract}
We present \emph{James Webb Space Telescope (JWST)} Near Infrared Spectrograph (NIRSpec) integral-field spectroscopy of the nearby luminous infrared galaxy, NGC~7469. 
We take advantage of the high spatial/spectral resolution and wavelength coverage of \emph{JWST}/NIRSpec to study the 3.3\um\ neutral polycyclic aromatic hydrocarbon (PAH) grain emission on $\sim60$~pc scales. We find a clear change in the average grain properties between the star-forming ring and the central AGN. Regions in the vicinity of the AGN, with \NeIII/\NeII$>$0.25, tend to have larger grain sizes and lower aliphatic-to-aromatic (3.4/3.3) ratios indicating that smaller grains are preferentially removed by photo-destruction in the vicinity of the AGN. We find an overall suppression of the total PAH emission relative to the ionized gas in the central 1\,kpc region of the AGN in NGC~7469 compared to what has been observed with Spitzer on 3 kpc scales. However, the fractional 3.3\um\ to total PAH power is enhanced in the starburst ring, possibly due to a variety of physical effects on sub-kpc scales, including recurrent fluorescence of small grains or multiple photon absorption by large grains. Finally, the IFU data show that while the 3.3\um\ PAH-derived star formation rate (SFR) in the ring is 8\% higher than that inferred from the [NeII] and [NeIII] emission lines, the integrated SFR derived from the 3.3\um\ feature would be underestimated by a factor of two due to the deficit of PAHs around the AGN, as might occur if a composite system like NGC~7469 were to be observed at high-redshift.

\end{abstract}



\section{Introduction} \label{sec:intro}

    



The strong infrared emission bands at 3.3, 6.2, 7.7, 8.6, 11.3, and 17\um\ are detectable in the spectra of a diverse range of astrophysical sources, ranging from photodissociation regions in nebulae \citep{Werner2004} to nearby star-forming galaxies \citep{Peeters2004a, Smith2007} and luminous infrared galaxies \citep[LIRGs; see][for a review]{Armus2020}, even to distant galaxies at high redshifts \citep[z$>$1;][]{Sajina2009, Pope2013, Riechers2014, Spilker2023}. These prominent spectral features are often attributed to Polycyclic aromatic hydrocarbon (PAH) molecules \citep{Leger1984, Allamandola1985}, which can account for up to 15\% of the cosmic carbon \citep{Li2001, Draine2007, Zubko2004, Jones2017} and contribute up to $\sim$20\% of the total infrared luminosity from a galaxy \citep{Smith2007}. 

The 3.3\um\ PAH band is the shortest wavelength feature among all the PAHs and is primarily attributed to the C-H stretching mode. It is known to trace the neutral, smallest PAH population in the interstellar medium (ISM), typically with a radius of $\sim$5\AA\ or N$_\mathrm{C}\sim$ 50 carbon atoms \citep{Schutte1993, Draine2007}. The 3.3\um\ PAH has been the least studied PAH feature due to its inaccessibility from the ground-based observations, and the lack of spectral coverage of Spitzer/IRS shortward of 5\um. Nonetheless, observations with the AKARI infrared satellite \citep{Murakami2007} had previously demonstrated the power of using the 3.3\um\ PAH to diagnose star formation and AGN in (U)LIRGs \citep{Imanishi2008, Imanishi2010, Ichikawa2014, Inami2018} and to study the photoelectric efficiency that couples the stellar radiation field with gas temperatures \citep{McKinney2021}. Just as importantly, the 3.3\um\ band is the only PAH feature that will be accessible to JWST at high-redshifts, making detailed studies of this feature at low-redshift of paramount importance for understanding dust and metals at early epochs.

In a joint AKARI+Spitzer study, \citet{Lai2020} found that the fractional 3.3\um\ PAH power (\Lpahthree/\Ltotpah) in nearby star-forming galaxies typically ranges from 1.5--3\%, a value that is significantly smaller than for the brightest 7.7\um\ PAH complex, which typically reaches $\mathrm{L_{PAH \ 7.7}}$/\Ltotpah $\sim$ 40\% in nearby star-forming galaxies \citep{Smith2007}. Despite its faintness, the 3.3\um\ PAH feature has been proposed as an accurate tool for measuring the average grain size when paired with the 11.3 and 17\um\ PAH bands, which also serve as tracers of neutral grains \citep[e.g.][]{Croiset2016, Maragkoudakis2020, Lai2020, Draine2021, Rigopoulou2021, Maragkoudakis2022, Sidhu2022, Maragkoudakis2023a, Maragkoudakis2023b}. The 3.3\um\ PAH holds the potential to be a sensitive diagnostic of ISM conditions, as well as an estimator of the star formation rate (SFR) \citep[e.g.][]{Peeters2004a, Shipley2016, Lai2020} accessible from nearby galaxies to high redshifts with JWST imaging and spectroscopy \citep{Inami2018, Evans2022, Leroy2023, Sandstrom2023}. These applications, however, strongly rely on our understanding of the properties of PAHs and how they depend on metallicity and react to the ambient radiation field, as PAHs can be faint or even absent in low metallicity environments \citep{Wu2006, Hao2009} as well as in galaxies with powerful AGN \citep{ODowd2009, Diamond-Stanic2010}. Surprisingly, some studies have shown that AGN photons can also be the source of PAH molecules excitation and emission \citep{Howell2007, Jensen2017}. 

In this paper, we focus on tracing the physical conditions of the neutral dust grains probed by the 3.3\um\ PAH in NGC~7469, using the superb spatial and spectral resolution of the \emph{JWST}/Near-InfraRed Spectrograph \citep[NIRSpec:][]{Jakobsen2022}. As a follow up study to \citet[][hereafter \paperI]{Lai2022}, the \emph{JWST}/MIRI results are also included here in a joint analysis, to provide a complete view of the aromatic bands in NGC~7469. 

NGC~7469 (Arp 298, Mrk 1514, IRAS F23007+0836) is a galaxy located at D$_{L}$=70.6 Mpc and is classified as a Seyfert 1.5 \citep{Landt2008}. It is a LIRG ($L_{8-1000 \um} = 10^{11.6} L_\sun$) with a supermassive black hole mass of 1.1$\times$10$^{7}$ M$_{\sun}$ \citep{Peterson2014, Lu2021} and X-ray luminosity of $L_\mathrm{2-10 keV} = 10^{43.19}$ erg s$^{-1}$ \citep{Asmus2015}. NGC~7469 contains an accreting black hole and a starburst ring with a radius of $\sim$500~pc \citep[e.g.][]{Song2021}. Recently, the NIRCAM imaging data revealed a factor of $\sim$6 more dusty, young ($<$5 Myr) clusters compared to the previous identifications made by HST \citep{Bohn2023}. Using the JWST integral field spectroscopy, emission lines with ionization potential up to 187 eV are detected within a distance of ~$\sim$100~pc from the AGN, with an outflowing wind that can reach up to 1700 km/s traced by the high ionization lines \citep{Armus2023}, and the ionized coronal wind probed by [Mg\,\textsc{v}] (IP: 109 eV) extends out to $\sim$400~pc from the nucleus \citep{U2022b}. By studying the PAH ratios in the mid-infrared, \paperI\ found that feedback from the central black hole mostly impacts the inner ISM region, leading to a larger grain size distribution in the vicinity of the AGN than that in the ring \citep[cf.][]{Garcia-Bernete2022b}. 



Throughout this paper, a cosmology with $H_{0}=70\,{\rm km\,s^{-1}\,Mpc^{-1}}$, $\Omega_{\rm M}=0.30$ and $\Omega_{\rm \Lambda}=0.70$ is adopted. The redshift of NGC~7469 ($z$=0.01627\footnote{NASA/IPAC Extragalactic Database (NED)}) corresponds to a projected physical scale of 330~pc per arcsecond.

\begin{figure*}
    \centering
        \includegraphics[width=0.495\textwidth]{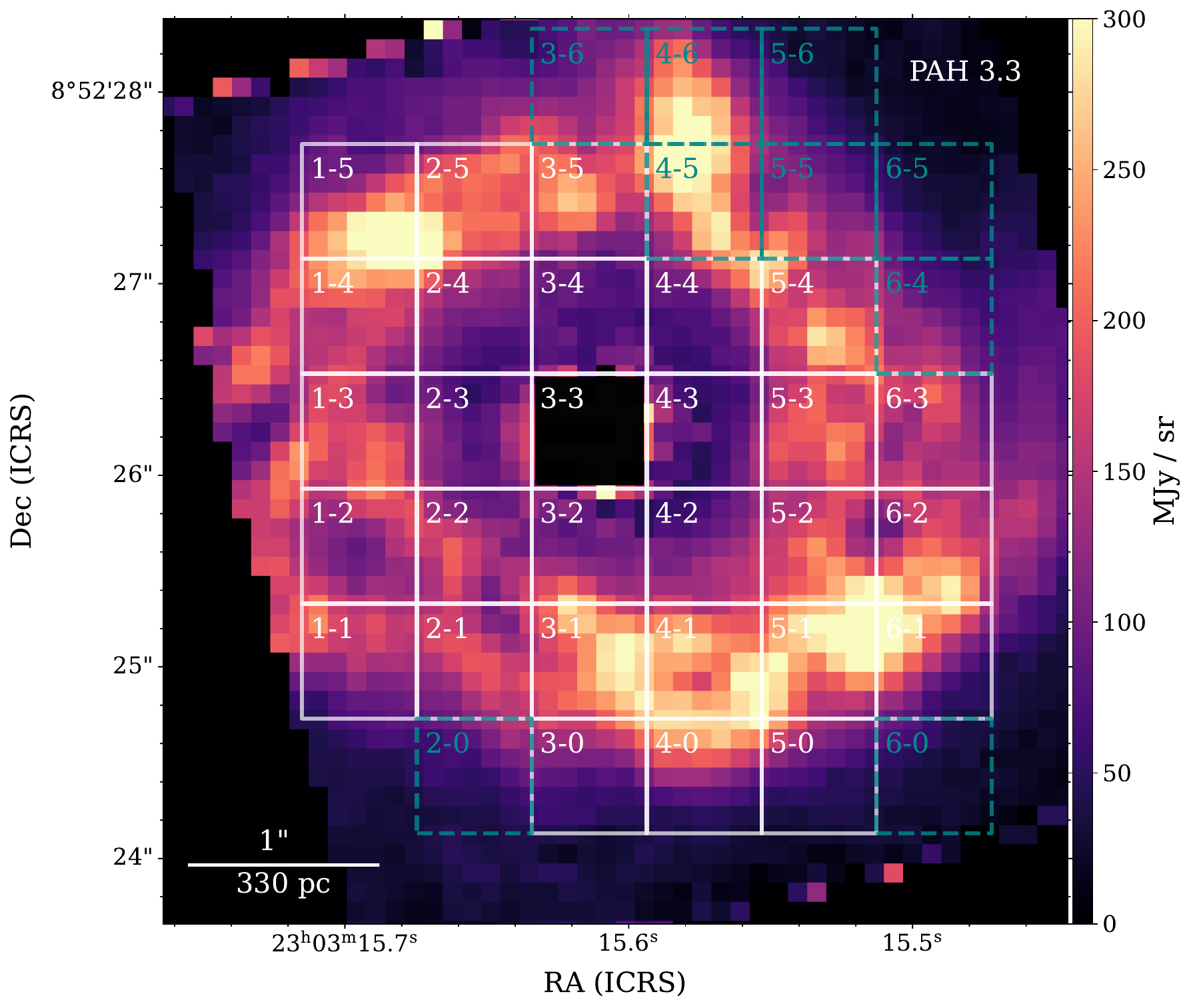} 
        \includegraphics[width=0.495\textwidth]{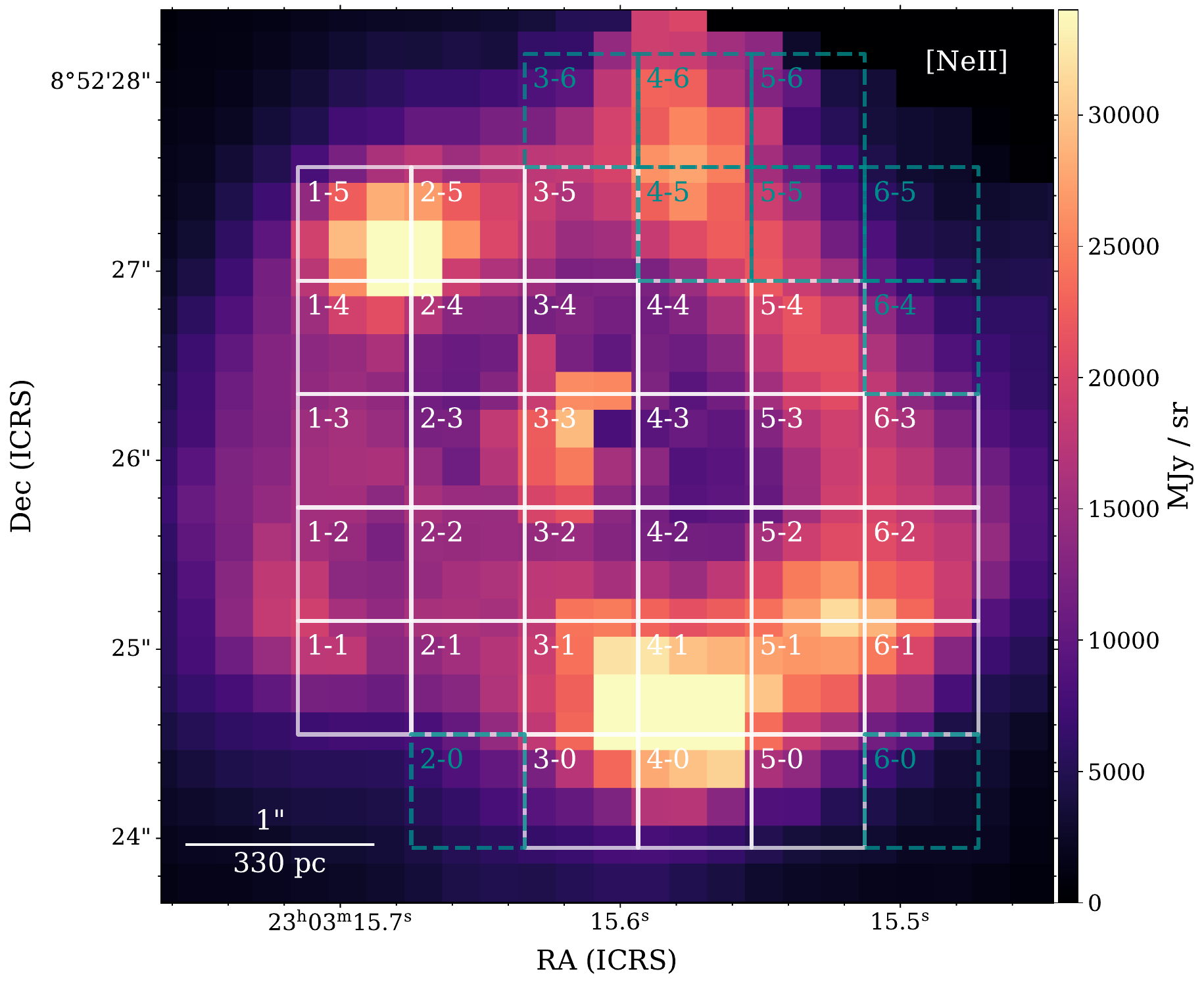} 
    \caption{(\textit{left}) The 3.3\um\ polycyclic aromatic hydrocarbon (PAH) map of NGC~7469. This image was created by subtracting a linear fit to the continuum in the unsmoothed NIRSpec cube (see \S\ref{sec:analysis}), and it has a pixel size of 0\farcs1, providing us with the highest spatial view of the dust distribution in NGC~7469 using JWST IFU. The spectra extracted in this study are based on the overlaid square grid with each cell having a width of 0\farcs6. Grids in white are cells with full NIRSpec (G395H)+MIRI coverage, while the ones in blue (dashed) have only partial coverage in MIRI. The central grid is masked due to noise in the linear continuum subtraction at the nucleus, where the PAH emission is very weak but has no effect on the extracted spectra analyzed in this paper.  
    (\textit{right}): The \NeII\ 12.81\um ~MIRI/MRS map for comparison, which has a pixel size of 0\farcs2, and traces the ionized atomic gas in the starburst ring. The same spectral grid is overlaid here. In general, there is good agreement between the 3.3 PAH and [NeII] maps, but the 3.3 PAH map shows finer detail owing to its higher intrinsic spatial resolution, and differences are evident in the ratio of PAH to ionized gas emission around the ring.} 
    \label{fig:pah33_image}
\end{figure*}



\begin{figure*}
 	\includegraphics[width=1\textwidth]{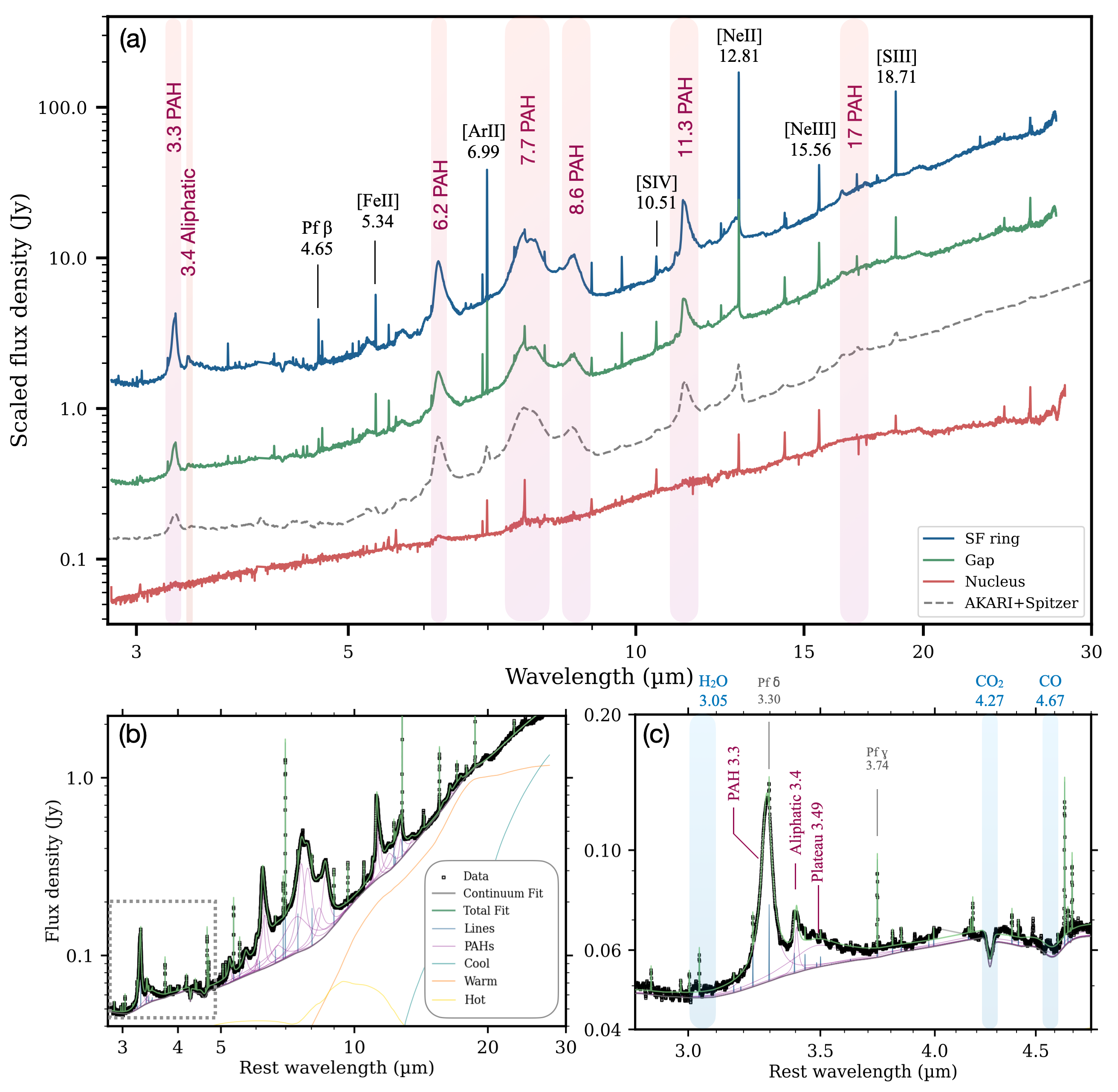}
    \caption{(a) The stacked (summed over all corresponding grid cells) spectra in the three regions of NGC 7469, including the star-forming ring (blue), the gap between the ring and the nucleus (green), and the nucleus (red), with the key diagnostic emission features labeled. The PAHs are heavily suppressed in the nucleus where the \NeIII/\NeII\ is highest, indicative of a harder radiation field dominated by the central AGN. The low-resolution (R$\sim$100) AKARI+Spitzer spectrum, which covers the nucleus, gap, and star-forming ring taken from the ASESS catalog \citep{Lai2020}, is also presented for comparison (dashed grey line). (b) shows the multi-component, CAFE fit of the stacked star-forming ring spectrum, with a zoom-in view of the 3\um\ regime in (c), delineated by the dashed box. As in (a), key features are labeled here. The aromatic feature at 3.3\um\ and the aliphatic feature at 3.4\um\ can be readily differentiated with the high spectral resolution of NIRSpec, allowing us for the first time to confidently separate these features in galaxies and use their flux ratio as an independent diagnostic of the dusty ISM.}
    \label{fig:Fig2}
\end{figure*}

\section{Observations and data reduction}
\label{sec:observation}


As part of the Director's Discretionary Time Early Release Science (ERS) program 1328 (Co-PIs: L. Armus and A. Evans), the \JWST\ Near-Infrared IFU observations \citep{Boker2022} on NGC~7469 were taken with NIRSpec on July 19th, 2022. The NIRSpec IFU observations were carried out using a set of high-resolution gratings (with a nominal resolving power of R$\sim$2700), namely G140H/100LP, G235H/170LP, and G395H/290LP, covering the wavelength of 0.97--5.27\um. For each grating, the science exposure time was 817 seconds, and a 4-pt dither pattern was used to sample the extended star-forming ring. In this study, we focus on the G395H grating that contains the shortest PAH band at 3.3\um. 

We downloaded the uncalibrated science and background observations through the MAST Portal. The data reduction process was done using the \JWST\ Science Calibration Pipeline version 1.8.3 \citep{Bushouse2022}, with a reference file of \texttt{jwst\_12027.pmap}. Three stages of the pipeline processing were applied, including \texttt{Detector1}, \texttt{Spec2}, and \texttt{Spec3}. While the dedicated background is not included in the observation strategy, ``leakcal" was performed in the observation to mitigate the contamination due to failed open micro-shutter assembly (MSA). We refer the readers to our companion paper (Bianchin et al. 2023; in prep.) for more details on the data processing. 

To obtain a complete spectroscopic view of NGC~7469, we combined the NIRSpec IFU data with the MIRI IFU data presented in \paperI, allowing us to study all the PAH features from 3 to 20\um. To properly compare the NIRSpec and MIRI data, two modifications to the NIRSpec cube had to be made. First, a spatial offset of $\sim$0\farcs24 between the NIRSpec and MIRI IFUs was found, so we combined the two IFUs by shifting the WCS of NIRSpec to match with that of MIRI, whose astrometry has been verified with the MIRI imaging. This offset was measured by comparing the collapsed images from the spectral overlap region between 5.0--5.1\um\ in both IFUs. Second, in order to combine the IFUs while matching their spatial resolution, we smoothed the NIRSpec G395H/290LP cube and MIRI channel~1 by convolving them with Gaussian kernels of 0.95 and 0.80 $\sigma$ (in pixel units), respectively, to match with the FWHM measured at 12\um, where the longest wavelength PAH feature (PAH 11.3\um\ in MIRI channel 2) studied in this \textit{Letter} is located.

\section{Analysis}
\label{sec:analysis}
Many studies have found that the hardness of the radiation field probed by \NeIII/\NeII\ plays an important role in regulating PAH ratios in AGN and star-forming galaxies \citep[e.g.][]{Smith2007, Sales2010, Garcia-Bernete2022a}. Particularly, in \paperI, we have shown that the main driver of grain size variation in this galaxy is \NeIII/\NeII, which will also be used here to understand the properties of the 3.3\um\ PAH. Thus, in Figure~\ref{fig:pah33_image} we present the \emph{unsmoothed} PAH and \NeII\ maps at 3.3\um\ and 12.81\um, respectively, both were created by applying a linear local continuum subtraction using the QFitsView tool \citep{Ott2012}. The star-forming ring is dominated by the bright PAH emission, which provides a dust map with the highest available spatial resolution of FWHM$\sim$0\farcs19 using the NIRSpec IFU data compared to the MIRI Ch3 \NeII\ map that has a FWHM$\sim$0\farcs48. In general, PAH 3.3 and \NeII\ emission show similar morphology, having four clumps along the ring, but the relative intensities differ. For example, the neon emission is brighter in the southern clump (4-1), whereas PAH 3.3 is brighter in the northern clump (4-5).

In \paperI, the study of the MIR dust and gas properties in NGC~7469 was carried out on an aperture-based analysis of the MIRI IFU data. In the study presented here, we maximize the full potential of an IFU observation by applying a grid extraction to both the NIRSpec and MIRI IFUs. This allows us to obtain spectra across a wide wavelength range from 3 to 28\um\ in a systematic approach. The grid extraction was performed on the smoothed NIRSpec and MIRI cubes using the spectral extraction tool included in the most recent version of the \textit{Continuum And Feature Extraction} \texttt{(CAFE)} software\footnote{https://github.com/GOALS-survey/CAFE}. This software was developed initially by \citet{Marshall2007} for \emph{Spitzer}/IRS and has recently been updated for \emph{JWST} (see D\'iaz-Santos et al.; in prep.), which includes both the capabilities of handling spectral extraction of the IFU cube and spectral decomposition of the extracted spectra. 

For the grid extraction, we initially set the center of the grid at the nucleus (23:03:15.6139, +8:52:26.230 J2000) and used 7$\times$7 grid cells, 0\farcs6 wide each, to extract spectra across the full field of view (FoV) from both the smoothed NIRSpec and MIRI cubes. The grid presented in Figure~\ref{fig:pah33_image} shows the positions of our extracted spectra. No extraction is done if the grid cell exceeds the coverage of the IFU FoV, so the total number of extraction is not 49 but rather 38, including the 29 cells having the full G395H+all MIRI sub-band coverage (white) and 9 cells having only the complete NIRSpec coverage but with an intermittent coverage in MIRI (dark blue). This partial coverage of MIRI is due to the slight spatial mismatch between each MIRI sub-channel. To obtain the full spectral coverage of 3---28\um, we stitched the spectrum of NIRSpec to MIRI. Among the 29 cells that cover the whole NIRSpec/G395H to MIRI range, we further divide them into three groups --- star-forming (SF), gap, and nucleus. Most of the extracted spectra belong to the star-forming ring except for the five cells, 2-3, 2-4, 3-4, 4-3, and 4-4, which are designated as gap region cells, and the nuclear extraction (3-3). Note that the nuclear spectrum presented here differs from the AGN spectrum analyzed in \citet{Armus2023}, which was extracted using an expanding cone with a small inner diameter of 0\farcs3 at 5.5\um\ to isolate the central source and the base of the outflow.

\section{Results}
\label{sec:Results}



The FoV of the dithered \JWST\ NIRSpec IFU observations (4\farcs2 $\times$ 4\farcs8) provides complete coverage of the NGC~7469 circumnuclear ring and inner ISM, allowing for high spatial and spectral resolution in the study of dust and star-formation properties in the near-infrared. To demonstrate the variation of the spectra across the IFU, we summed the individual grid spectra to create three NGC 7469 spectra:  a nuclear, ring, and gap spectrum (see Figure~\ref{fig:Fig2}(a)).  The low-resolution (R$\sim$100) AKARI+Spitzer spectrum taken from the AKARI-Spitzer Extragalactic Spectral Survey \citep[ASESS;][]{Lai2020} is also presented for comparison. Clearly, PAHs are heavily suppressed in the nuclear spectrum, while the high ionization fine structure lines are noticeably strong.  For example, the nuclear spectrum has the highest \NeIII/\NeII\ and \SIV/\NeII\ line flux ratios among the region spectra.

To fit and analyze the NIRSpec spectra, we use \texttt{CAFE}, a spectral decomposition tool that is able to simultaneously fits the PAH features, the dust continuum, the absorbed ice and gas features, and the narrow fine structure atomic and molecular gas emission lines. The stacked star-forming ring spectrum together with its spectral fit is presented in Fig.~\ref{fig:Fig2}(b) and a zoom-in view at the 3\um\ region is presented in Fig.~\ref{fig:Fig2}(c). In this regime, the spectrum is dominated by the 3.3\um\ PAH band and the 3.4\um\ aliphatic feature, sitting on top of a broad plateau that can extend up to $\sim$3.8\um. In addition, absorption features due to water ice at 3.05\um, CO$_{2}$ ice at 4.27\um, and CO ice at 4.67\um\ can also be seen. We report the central wavelengths and widths of the three dust components seen in the stacked ring spectrum in Table~\ref{tab:dust_param}. We find the 3.3 and 3.4\um\ features to be narrower, while the plateau is broader than previous measurements made with AKARI \citep{Lai2020}. This is undoubtedly due to the substantially higher spectral resolution of NIRSpec, enabling an accurate and detailed deblending of the dust components in the 3\um\ complex for the first time. The CAFE measurements of the extinction-corrected PAH and neon fluxes in each cell are presented in Table~\ref{tab:flux}. 

\begin{deluxetable}{@{\extracolsep{40pt}}lcc}
\tabletypesize{\footnotesize}
\tablewidth{0pt}
 \tablecaption{The 3\um\ dust features }\label{tab:dust_param}
 \tablehead{
 \colhead{Component} &  \colhead{$\lambda$ [\um]} & \colhead{$\Gamma$}\\ [-0.2cm]
 \colhead{(1)} & \colhead{(2)} & \colhead{(3)}\\ [-0.5cm]
 }
 \startdata 
3.3 PAH & 3.288 &  0.012\\
3.4 aliphatic & 3.404 & 0.005\\
Plateau & 3.484 &  0.070\\
 \enddata
 \vspace{0.1cm}
 \tablecomments{Central wavelength and widths of the dust feature as measured in the stacked NGC~7469 star-forming ring spectrum. Col. (1): dust feature name; (2): central wavelength of the feature; (3): fractional FWHM, defined by the ratio of FWHM and central wavelength.}
\end{deluxetable}

\begin{figure*}
    \centering
        \includegraphics[width=0.495\textwidth]{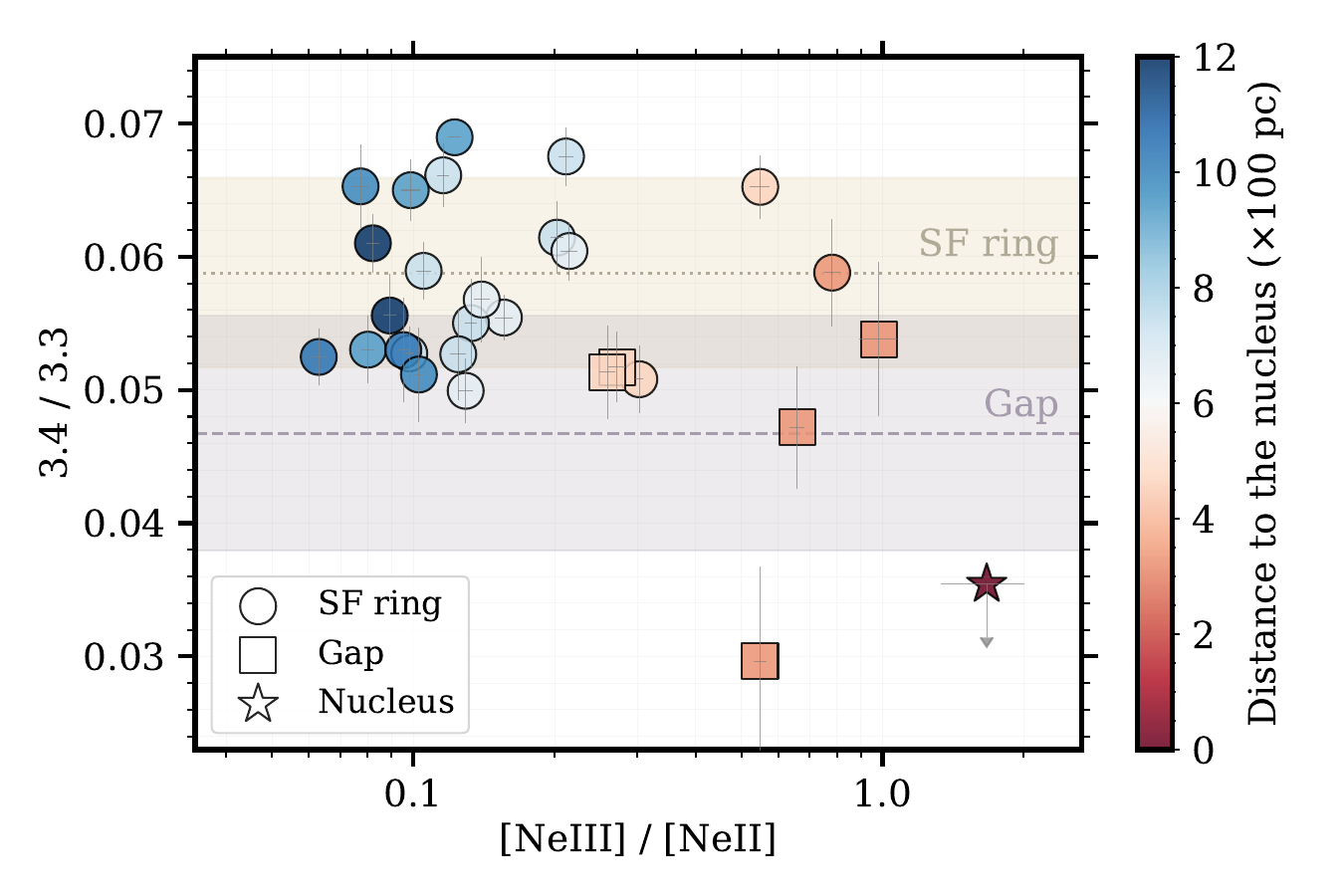} 
        \includegraphics[width=0.495\textwidth]{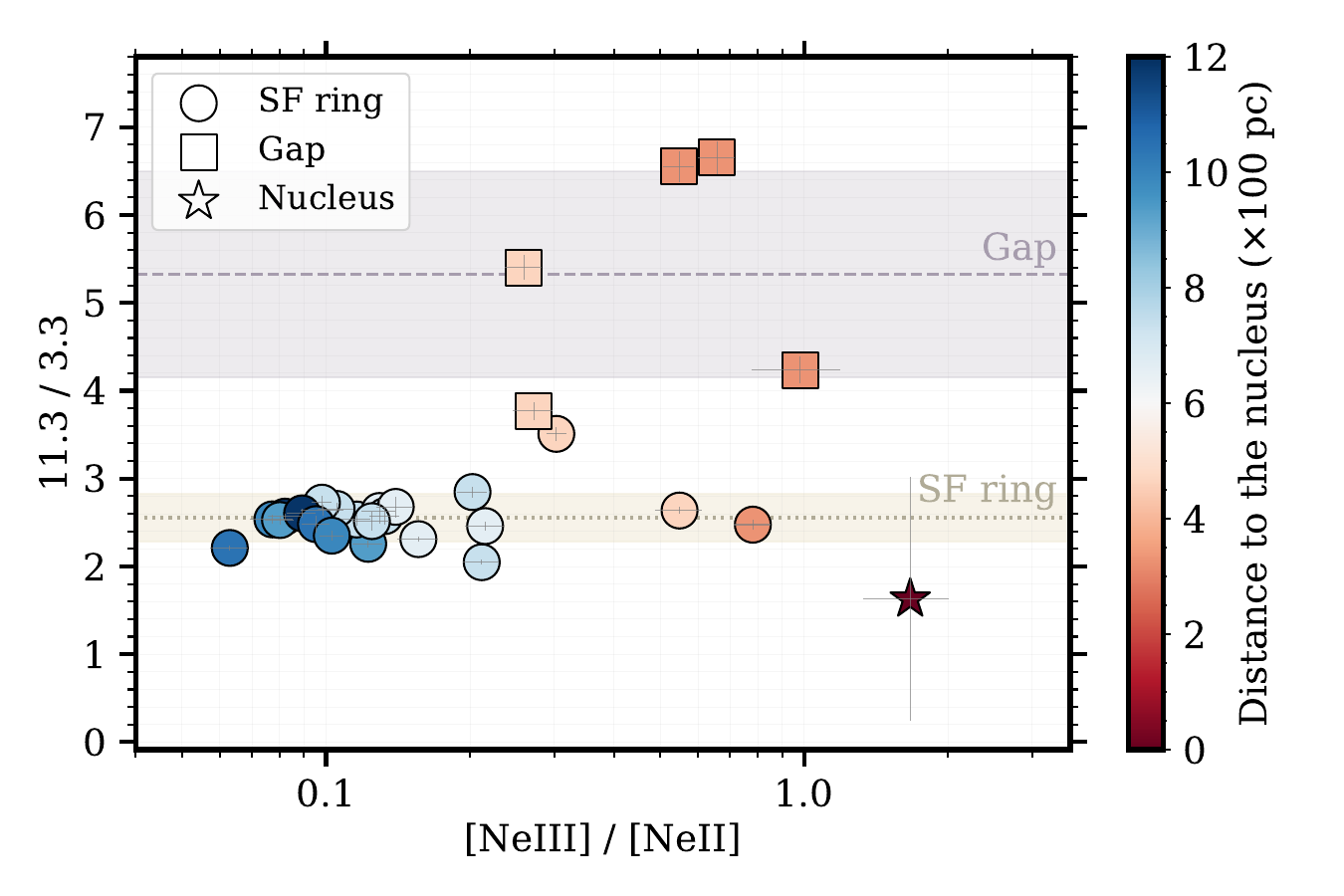} 
    \caption{(\textit{left}): The aliphatic (3.4\um) to aromatic (3.3\um) ratio as a function of the hardness of the radiation field probed by \NeIII/\NeII\ in NGC 7469. The circles indicate cells in the star-forming ring, the squares indicate cells in the gap region, and the star indicates the nucleus. The color indicates the projected linear distance from the AGN. The two dashed lines and the shaded regions indicate the averages and 1-sigma of 3.4/3.3 in the gap region cells (squares) and the star-forming ring (circles). On average, the 3.4/3.3 ratio in the gap decreases by 20\% compared to that of the star-forming ring. The 3.4/3.3 ratio shows a large scatter when \NeIII/\NeII$>$0.5. The 3.4\um\ feature at the nucleus is not detected; a 3-sigma upper limit is shown here. (\textit{right}): The 11.3/3.3 PAH ratio (a diagnostic of the average sizes of the PAH grains) as a function of the hardness of the radiation field. The two dashed lines and the shaded regions indicate the averages and 1-sigma of PAH 11.3/3.3 in the gap region cells (squares) and the star-forming ring (circles). Cells in the gap region show on average a factor of 2 higher 11.3/3.3 compared to those in the star-forming ring, suggesting a larger average PAH grain size in regions with a harder radiation field.  This may be due to the preferential destruction of small grains by the AGN. The nucleus appears to show an anomalously low 11.3/3.3 ratio but with relatively large uncertainty.}
    \label{fig:pah33_ali34_Ne_image}
\end{figure*}

\begin{deluxetable*}{@{\extracolsep{4pt}}ccrrrrrrr}
\tabletypesize{\footnotesize}
\tablewidth{0pt}
 \tablecaption{Extinction-corrected PAH Band and Emission Line Fluxes in NGC~7469}\label{tab:flux}
 \tablehead{
 \colhead{Cell} &  \colhead{Group} & \colhead{PAH 3.3\um} & \colhead{Aliphatic 3.4\um} & \colhead{PAH 11.3\um} & \colhead{$\Sigma$PAH} & \colhead{[NeII] 12.81\um} & \colhead{[NeIII] 15.56\um} \\ [-0.2cm]
  \colhead{} & & \colhead{[$\times$10$^{-17}$ W/m$^{2}$]} & \colhead{[$\times$10$^{-17}$ W/m$^{2}$]} & \colhead{[$\times$10$^{-17}$ W/m$^{2}$]} & \colhead{[$\times$10$^{-17}$ W/m$^{2}$]} & \colhead{[$\times$10$^{-17}$ W/m$^{2}$]} & \colhead{[$\times$10$^{-17}$ W/m$^{2}$]} \\ [-0.4cm]
 }
 \startdata 
1-1 & SF &  6.20 $\pm$ 0.01 &  0.43 $\pm$ 0.01 &   14.00 $\pm$  0.25 &   101.02 $\pm$  1.07 &  3.82 $\pm$ 0.14 &   0.47 $\pm$  0.03 \\
1-2 & SF &  5.43 $\pm$ 0.04 &  0.33 $\pm$ 0.01 &   15.45 $\pm$  0.31 &   118.35 $\pm$  2.27 &  3.63 $\pm$ 0.18 &   0.74 $\pm$  0.04 \\
1-3 & SF &  5.42 $\pm$ 0.03 &  0.33 $\pm$ 0.01 &   13.32 $\pm$  0.27 &   120.55 $\pm$  1.44 &  3.47 $\pm$ 0.22 &   0.75 $\pm$  0.05 \\
1-4 & SF &  7.05 $\pm$ 0.04 &  0.47 $\pm$ 0.02 &   17.86 $\pm$  0.46 &   149.37 $\pm$  2.05 &  5.46 $\pm$ 0.17 &   0.63 $\pm$  0.04 \\
1-5 & SF &  4.16 $\pm$ 0.05 &  0.27 $\pm$ 0.01 &   10.97 $\pm$  0.22 &    86.18 $\pm$  1.34 &  3.98 $\pm$ 0.22 &   0.39 $\pm$  0.03 \\
2-1 & SF &  7.24 $\pm$ 0.04 &  0.49 $\pm$ 0.02 &   14.83 $\pm$  0.18 &   109.46 $\pm$  1.18 &  3.75 $\pm$ 0.25 &   0.79 $\pm$  0.04 \\
2-2 & SF &  8.29 $\pm$ 0.09 &  0.54 $\pm$ 0.02 &   21.85 $\pm$  0.24 &   157.93 $\pm$  2.20 &  4.69 $\pm$ 0.31 &   2.57 $\pm$  0.24 \\
2-3 & G &  2.27 $\pm$ 0.06 &  0.12 $\pm$ 0.01 &    9.60 $\pm$  0.24 &    80.87 $\pm$  1.21 &  2.90 $\pm$ 0.41 &   2.85 $\pm$  0.44 \\
2-4 & G &  4.61 $\pm$ 0.10 &  0.24 $\pm$ 0.01 &   17.36 $\pm$  0.29 &   128.20 $\pm$  2.36 &  5.27 $\pm$ 0.40 &   1.43 $\pm$  0.09 \\
2-5 & SF &  5.98 $\pm$ 0.10 &  0.35 $\pm$ 0.01 &   15.89 $\pm$  0.31 &   128.15 $\pm$  1.51 &  5.26 $\pm$ 0.37 &   0.55 $\pm$  0.03 \\
3-0 & SF &  5.72 $\pm$ 0.10 &  0.37 $\pm$ 0.02 &   14.50 $\pm$  0.16 &    99.16 $\pm$  1.37 &  5.97 $\pm$ 0.38 &   0.46 $\pm$  0.04 \\
3-1 & SF &  8.75 $\pm$ 0.04 &  0.49 $\pm$ 0.01 &   20.22 $\pm$  0.26 &   167.26 $\pm$  3.20 &  6.51 $\pm$ 0.57 &   1.02 $\pm$  0.05 \\
3-2 & SF & 10.23 $\pm$ 0.18 &  0.60 $\pm$ 0.04 &   25.32 $\pm$  0.41 &   154.65 $\pm$  3.77 &  5.34 $\pm$ 0.42 &   4.17 $\pm$  0.18 \\
3-3 & N &  3.24 $\pm$ 1.30 &  0.04 $\pm$ 0.53 &    5.28 $\pm$  3.96 &    98.51 $\pm$  4.17 &  3.77 $\pm$ 0.75 &   6.28 $\pm$  0.24 \\
3-4 & G &  1.71 $\pm$ 0.04 &  0.05 $\pm$ 0.01 &   11.21 $\pm$  0.17 &    77.14 $\pm$  6.72 &  2.91 $\pm$ 0.17 &   1.60 $\pm$  0.07 \\
3-5 & SF &  5.74 $\pm$ 0.11 &  0.29 $\pm$ 0.01 &   15.10 $\pm$  0.75 &   128.57 $\pm$  2.27 &  4.36 $\pm$ 0.21 &   0.56 $\pm$  0.04 \\
4-0 & SF &  7.77 $\pm$ 0.04 &  0.41 $\pm$ 0.02 &   17.18 $\pm$  0.22 &   125.52 $\pm$  1.52 &  9.21 $\pm$ 0.44 &   0.58 $\pm$  0.05 \\
4-1 & SF &  7.46 $\pm$ 0.12 &  0.39 $\pm$ 0.01 &   20.34 $\pm$  0.52 &   166.68 $\pm$  2.59 &  7.22 $\pm$ 0.27 &   0.71 $\pm$  0.02 \\
4-2 & SF & 10.88 $\pm$ 0.21 &  0.55 $\pm$ 0.03 &   38.17 $\pm$  0.48 &   279.09 $\pm$  3.69 &  8.58 $\pm$ 0.27 &   2.60 $\pm$  0.09 \\
4-3 & G &  3.15 $\pm$ 0.05 &  0.15 $\pm$ 0.01 &   20.99 $\pm$  0.47 &   150.66 $\pm$  4.55 &  4.55 $\pm$ 0.38 &   2.99 $\pm$  0.09 \\
4-4 & G &  2.53 $\pm$ 0.06 &  0.13 $\pm$ 0.01 &   13.68 $\pm$  0.15 &    89.45 $\pm$  1.09 &  4.05 $\pm$ 0.23 &   1.05 $\pm$  0.07 \\
5-0 & SF &  6.74 $\pm$ 0.04 &  0.41 $\pm$ 0.01 &   17.31 $\pm$  0.23 &   110.81 $\pm$  5.84 &  6.22 $\pm$ 0.25 &   0.51 $\pm$  0.03 \\
5-1 & SF &  8.79 $\pm$ 0.06 &  0.47 $\pm$ 0.02 &   22.22 $\pm$  0.40 &   182.85 $\pm$  1.91 &  7.42 $\pm$ 0.29 &   0.59 $\pm$  0.03 \\
5-2 & SF &  6.91 $\pm$ 0.19 &  0.38 $\pm$ 0.02 &   17.78 $\pm$  0.47 &   151.58 $\pm$  1.93 &  5.18 $\pm$ 0.21 &   0.69 $\pm$  0.04 \\
5-3 & SF &  6.34 $\pm$ 0.15 &  0.36 $\pm$ 0.02 &   16.97 $\pm$  0.62 &   146.05 $\pm$  2.67 &  5.16 $\pm$ 0.34 &   0.72 $\pm$  0.04 \\
5-4 & SF &  5.85 $\pm$ 0.13 &  0.31 $\pm$ 0.02 &   14.72 $\pm$  0.55 &   122.43 $\pm$  1.97 &  4.94 $\pm$ 0.29 &   0.62 $\pm$  0.04 \\
6-1 & SF &  8.02 $\pm$ 0.06 &  0.45 $\pm$ 0.03 &   20.89 $\pm$  0.35 &   168.21 $\pm$  1.63 &  6.29 $\pm$ 0.26 &   0.56 $\pm$  0.03 \\
6-2 & SF &  6.74 $\pm$ 0.23 &  0.36 $\pm$ 0.02 &   16.72 $\pm$  0.32 &   141.98 $\pm$  1.66 &  5.49 $\pm$ 0.15 &   0.52 $\pm$  0.03 \\
6-3 & SF &  6.57 $\pm$ 0.11 &  0.34 $\pm$ 0.02 &   15.44 $\pm$  0.29 &   115.76 $\pm$  2.71 &  4.57 $\pm$ 0.13 &   0.47 $\pm$  0.02 \\
 \enddata
 \vspace{0.1cm}
 \tablecomments{In the column Group, ``SF'' refers to the star-forming ring cell, ``G'' refers to the Gap cell, and ``N'' refers to the nucleus cell.}
\end{deluxetable*}

\subsection{Aliphatic versus aromatic emission in NGC~7469}
\label{sect:aiphatic_aromatic}
The abundance ratios of different types of carbonaceous dust can provide insights into their formation mechanisms and histories. The 3.4\um\ emission feature is often attributed to the vibrations of -CH3 (methyl) and -CH2- (methylene) groups in aliphatic hydrocarbons \citep{Joblin1996, Yang2016}, and it can also be seen in absorption along sight lines with large extinction \citep{Pendleton2002, Chiar2013, Hensley2020}.  The variation of the ratio between this 3.4\um\ aliphatic feature and the 3.3\um\ aromatic feature is indicative of the processing of dust particles in the ISM, as the chain-like aliphatic bonds are more fragile compared to the ring-like aromatic bonds. 

The 3.4\um\ aliphatic feature was detected in all the grid cells of NGC 7469 except for the nucleus. The 3.4/3.3 ratio varies by about a factor of two across the IFU FoV. Regions with a harder radiation field, as measured by the \NeIII/\NeII\ line flux ratio tend to have a lower average 3.4/3.3 ratio and a larger scatter. The 3.4/3.3 ratio is lower in cells between the ring and the nucleus, decreasing by about 20\% compared to those in the star-forming ring, suggesting that the aliphatic bonds are more susceptible to photo-destruction than the aromatic bonds (Figure~\ref{fig:pah33_ali34_Ne_image} (left)). There is no detection of the 3.4\um\ emission at the nucleus, and the range in 3.4/3.3 ratio values is largest in the cells with \NeIII/\NeII$>$0.5. While visible, the variation of the 3.4/3.3 ratio in the star-forming ring of NGC~7469 is relatively small, ranging only from 5---7\%, compared to Galactic regions that can range widely from $\sim$1---17\% in PDRs \citep[][see Figure 7(c)]{Pilleri2015}, suggesting the photo energy density in the ring is high, with log(G$_\mathrm{0}$)$>$3 in units of the Habing field \citep{Habing1968}.


\subsection{PAH Size Distribution}
\label{sect:113_33}
Small particles like PAHs tend to have larger cross-sections per unit mass than larger grains when assuming a continuous grain size distribution \citep{Mathis1977}, rendering them sensitive probes for the local physical conditions such as size and charge state within the ISM. Smaller PAHs tend to emit at short wavelengths because upon absorbing a UV photon, each bond in a smaller PAH molecule gets excited to a higher vibrational level, which will subsequently emit in short wavelengths. Typically, PAH size estimates have relied on the ratio of 6.2/7.7 PAH, while the 11.3/7.7 PAH ratio has been used for determining charge states \citep{Draine2001}. The ratio of 6.2/7.7, however, is not an ideal tracer for the size because the difference between the two wavelengths is relatively small, so the ratio has only limited diagnostic power as shown in Fig. 7 of \citet{Draine2007}. 

One of the key advantages of NIRSpec is accessing the 3.3\um\ PAH band, which has been shown to be a more effective proxy for the size of the small dust populations, particularly when used together with the 11.3\um\ band, since both trace neutral PAHs \citep[e.g.][]{Croiset2016, Maragkoudakis2020, Sidhu2022}. Figure~\ref{fig:pah33_ali34_Ne_image} (right) shows the 11.3/3.3 PAH ratio as a function of the hardness of the radiation field. In the star-forming spectra of NGC~7469, the PAH 11.3/3.3 ratios cluster around a value of 2.6$\pm$0.3, with a relatively smaller scatter. The gap spectra, on the other hand, exhibit on average a 2$\times$ increase in the 11.3/3.3 PAH ratios with a much broader distribution (5.3$\pm$1.2). This agrees with the findings presented in \paperI, which suggested that the average grain size becomes larger closer to the nucleus, consistent with the destruction of small grains in the harder UV fields closer to the AGN. The nucleus appears to display an anomalously low 11.3/3.3 ratio but with a relatively large uncertainty due to very weak PAH emission (see the nucleus spectrum in Figure~\ref{fig:Fig2}(a)).

\begin{figure}
    \centering
        \includegraphics[width=0.5\textwidth]{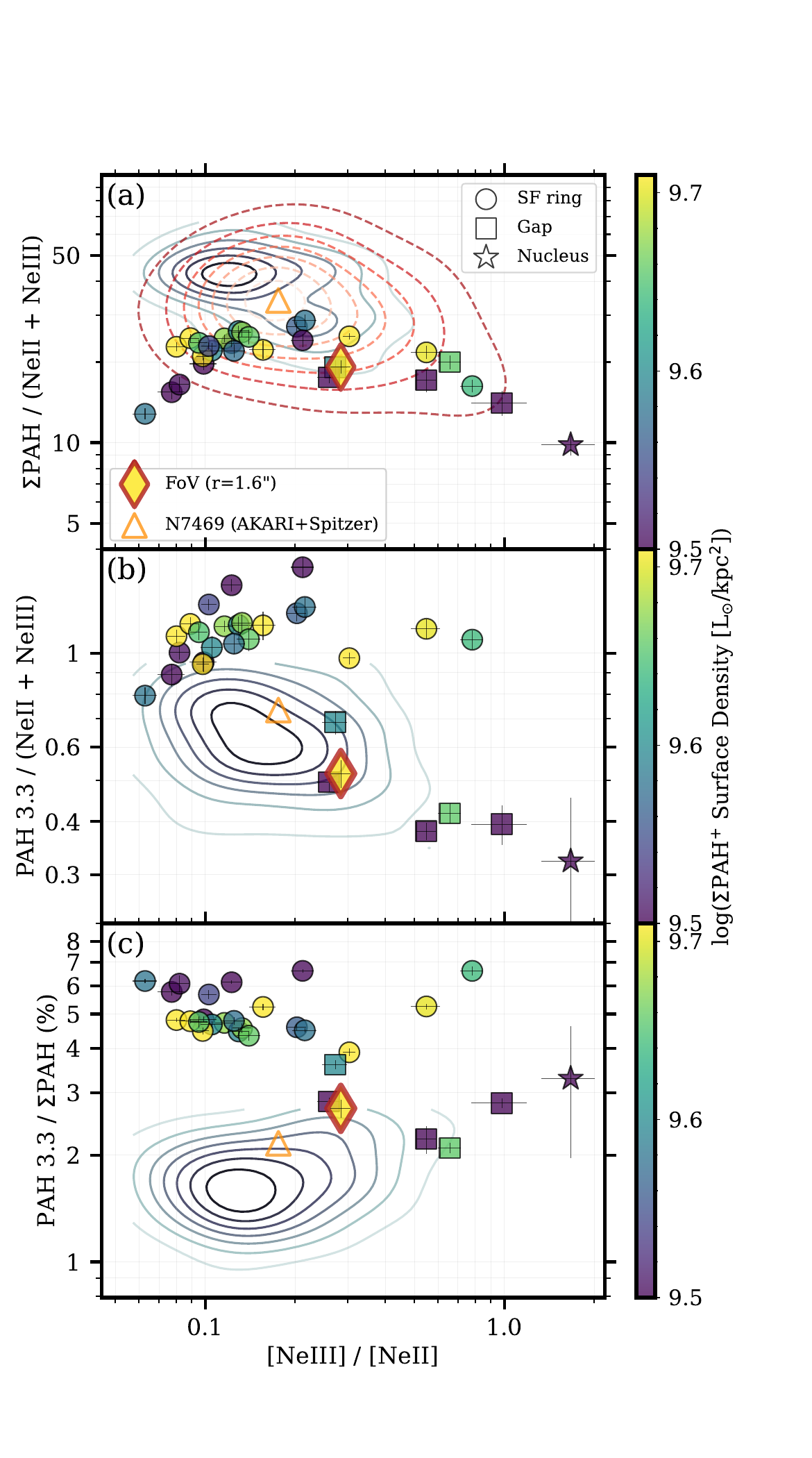} 
    \caption{Fractional PAH emission as a function of radiation field hardness in NGC~7469. The data are color-coded by the surface density of the total ionized PAH, consisting of the 6.2, 7.7, and 8.6\um\ PAH bands, which traces the amount of photoelectric heating in the ISM. Symbols are as in Figure~\ref{fig:pah33_ali34_Ne_image}, with the addition of the red diamond representing values measured from the total IFU spectrum, and the orange triangle representing values measured from the combined AKARI+Spitzer spectrum. For comparison, the red dashed contours indicate the distribution of LIRGs taken from \citet{Stierwalt2014}, and the blue solid contours indicate the distribution of the 62 PAH-bright galaxies (mostly LIRGs that are also in GOALS) taken from \citet{Lai2020} but re-fitted with CAFE. While the $\Sigma$PAH-to-Ne ratio in NGC~7469 (\textit{top}) is consistent with the lower envelope of LIRG values, the 3.3-to-Ne ratio (\textit{middle}) is high, consistent with a relative excess of 3.3 PAH emission compared to other PAH features (\textit{bottom}), where the values in the ring of NGC~7469 are up to 4$\times$ higher than in most starbursting LIRGs (see \S\ref{sec:enhancement} and \S\ref{sec:discussion_excess_PAH33} for more detailed discussions).
    } 
    \label{fig:pah33_totPAH_Ne}
\end{figure}

\subsection{Enhancement of the 3.3 micron PAH}
\label{sec:enhancement}
In \paperI, it was shown that the hardness of the radiation field was correlated with the PAH size variations seen in NGC~7469. Here we expand that analysis using the 3.3\um\ PAH feature. Figure~\ref{fig:pah33_totPAH_Ne}(a) shows the variation of the total PAH-to-[NeII]+[NeIII] ratio as a function of the \NeIII/\NeII\ line flux ratio, color-coded by the total ionized PAH surface density, which traces the amount of photoelectric heating in the ISM. Here, the total PAH ($\Sigma$PAH) refers to the sum of the 3.3, 6.2, 7.7, 8.6, 11.3, and 17\um\ PAH features, while the total ionized PAH surface density is derived from only the 6.2, 7.7, and 8.6\um\ PAH features. The cells with \NeIII/\NeII $<$ 0.25 show an increasing PAH to Ne ratio with radiation field hardness. This trend is especially evident in the three cells at the lower left with very low PAH-to-Ne ratios, namely 3-0, 4-0, and 5-0, which are coincident with the youngest clusters ($<$5Myr) in the star-forming ring \citep{Diaz-Santos2007}. Their low $\Sigma$PAH-to-Ne ratios may be due to excess \HII\ region emission from hot stars together with a possible deficit of total PAH emission, similar to what has recently been observed in four nearby star-forming galaxies with JWST and MUSE \citep{Egorov2023}. Note that these cells have \NeIII/\NeII\ ratios that lie well within the observed range of (extra)galactic \HII\ regions \citep[e.g., 0.04---10.00 in][]{Martin-Hernandez2002}.  

In cells with large \NeIII/\NeII\ line flux ratios ($>$\,0.25), the $\Sigma$PAH-to-Ne ratio again drops, resulting in a downward trend of $\Sigma$PAH-to-Ne with increasing radiation hardness. Besides the grid cell spectral extractions, we include two additional points to Figure~\ref{fig:pah33_totPAH_Ne}. The diamond symbol represents the total extraction of the IFU using a radius of 1\farcs6 that covers the AGN and the entire ring, while the triangle represents the integrated light obtained from the combined AKARI+Spitzer NGC~7469 spectrum in the ASESS sample \citep{Lai2020}. The AKARI+Spitzer measurement probes an area that is three times wider than the coverage provided by the IFU, with a measured $\Sigma$PAH-to-Ne ratio that is larger than all the individual JWST points. This is consistent with \citet{Diaz-Santos2011} who showed that PAHs are more extended on average than the \NeII\ emission in the GOALS galaxies. The two sets of contours indicate the distributions based on 244 galaxies within GOALS \citep[][red]{Stierwalt2014} and 62 galaxies that are both in GOALS and the bright PAH samples in ASESS \citep[blue;][]{Lai2020} but re-fitted with CAFE to rule out the potential discrepancy led by the use of different spectral decomposition tools. The GOALS sample includes LIRGs powered by starbursts and/or AGN, while the bright PAH sample consists only of star-forming galaxies. The NGC~7469 data points overlap with the two sets of contours but lie at the lower envelope of both distributions, suggesting slightly weaker overall PAH emission (with respect to fine structure line emission) in NGC~7469 compared to global measurements of other LIRGs and star-forming galaxies. We note that only PAHs at wavelengths $>$5\um\ are considered in \citet{Stierwalt2014}, so there is no inclusion of the 3.3\um\ PAH, which typically contributes less than 3\% of the total PAH emission in star-forming galaxies \citep{Lai2020} but would nevertheless raise the GOALS points even further. 

In Figure~\ref{fig:pah33_totPAH_Ne}(b), the PAH 3.3-to-Ne ratios show an increasing trend in the SF ring when \NeIII/\NeII$<$0.25, suggesting small PAHs are more efficiently excited as the hardness of the radiation field increases, but as the \NeIII/\NeII\ ratio becomes larger ($>$ 0.25), the PAH 3.3/neon ratio drops by at least 50\%, most clearly seen in the gap cells between the nucleus and the ring. The relative enhancement of the 3.3\um\ PAH emission compared to other PAH features is evident in Figure~\ref{fig:pah33_totPAH_Ne}(c). Here, a striking picture emerges when we zoom in on the scale of the SF ring, with each cell having a width of $\sim$200~pc, and find PAH 3.3/$\Sigma$PAH ratios that can be up to 4$\times$ higher than those global measurements seen in other starburst galaxies. Another PAH band exhibiting a noteworthy enhancement is the 6.2\um\ feature; however, its maximum increase is limited to 30\%, a magnitude that falls significantly short of the enhancement observed in the 3.3\um\ PAH band.

\section{Discussion}
\label{sect:discussion}

\subsection{The excess of PAH 3.3}
\label{sec:discussion_excess_PAH33}
In \S~\ref{sec:Results}, we showed that the hardness of the radiation field plays a pivotal role in shaping the properties of the small dust grains responsible for PAH emission in NGC~7469, particularly the aliphatic-to-aromatic ratio and the grain sizes. The typical ratio of 3.4/3.3 in NGC~7469 is 3--7\% (Figure~\ref{fig:pah33_ali34_Ne_image}(a)) with the lowest values seen where 0.5$<$\NeIII/\NeII$<$1. The 3.4/3.3 has been shown to increase up to $\sim$20\% in some PDR environments such as the reflection nebula NGC~7023 \citep{Pilleri2015} and normal star-forming galaxies \citep{Lai2020}. The decrease of the 3.4/3.3 seen in the gap between the nucleus and the SF ring of NGC~7469 indicates further grain processing within 500~pc of the AGN. The nucleus itself (r$<$100~pc) shows no detectable aliphatic emission. 

The apparent relative excess of the 3.3\um\ PAH in NGC~7469 compared to star-forming galaxies is surprising, since the 3.3\um\ PAH emission traces the smallest PAH population in the ISM \citep[N$_\mathrm{C}\sim$ 50;][]{Draine2021}, which is thought to be most vulnerable to photon-destruction. In fact, all the PAHs are suppressed in NGC~7469 (Figure~\ref{fig:pah33_totPAH_Ne}(a)) compared to galaxies powered by starbursts, while the 3.3\um\ PAH is not suppressed as much as the other PAHs in the SF ring (Figure~\ref{fig:pah33_totPAH_Ne}(b)). This leads to the relative enhancement in the 3.3\um\ PAH flux. The 3.3\um\ PAH does exhibit a slight positive correlation with radiation hardness, dropping at the highest values of the \NeIII/\NeII\ ratio. The enhancement of the 3.3\um\ PAH emission in NGC~7469 can be seen even more clearly when studying the fractional PAH 3.3 in the ring, which can be $\sim$3---4$\times$ higher compared to starburst galaxies observed with AKARI+Spitzer (Figure~\ref{fig:pah33_totPAH_Ne}(c)).  Even in the gap regions, where the PAHs are clearly affected by the AGN, the fractional 3.3\um\ PAH power still lies at the upper end of the galaxy distribution. This enhancement of the 3.3\um\ PAH emission is consistent with other recent findings based on the JWST NIRCam and MIRI imaging of normal star-forming galaxies that seem to suggest an increase of the 3.3\um\ PAH's contribution in the F335M filter in regions with harder and more intense radiation fields \citep{Chastenet2023}. Our result also suggests that those cells in the star-forming ring having higher fractional 3.3\um\ PAH emission also have a lower total ionized PAH surface density, which is not surprising as 3.3\um\ PAH emission mainly traces neutral PAHs. 

When comparing our JWST IFU to Spitzer results, it is important to realize that the physical areas covered by our single cell in the JWST IFU and the Spitzer slit in NGC~7469 differ substantially, with the cell probing a region of 200~pc versus the IRS slit, which summed the light over $\sim$3~kpc scales. Based on Figure~\ref{fig:pah33_totPAH_Ne} (a) and (c), the total PAH is more suppressed in the central 1~kpc region (diamond; extracted by an aperture with r\,=\,1\farcs6) compared to the more coarse 3~kpc area (triangle), whereas the fractional 3.3\um\ PAH is more enhanced. We have ruled out the possibility of systematic differences in the flux estimates by \texttt{CAFE} when analyzing spectra with diverse spectral resolutions (R=100 vs. R=2700) to be the cause of such a PAH 3.3 enhancement.  A test has been done to smooth the high-resolution JWST spectrum of the total IFU extraction to match with the Spitzer's, and the ratios of 3.3/$\Sigma$PAH show only a negligible difference ($<$2\%).



Understanding the physics behind the enhanced durability of the smallest dust particles, which give rise to the excess emission from PAH 3.3 emitters, carries significant implications for simulating the dust mass build-up and chemical enrichment throughout cosmic time \citep[e.g.,][]{Narayanan2023}. Theoretical studies have shown that molecules consisting of fewer carbon atoms (N$_\textrm{c}\lesssim$40) are able to efficiently dissipate the absorbed UV energy via recurrent fluorescence (RF), with a relaxation timescale of $\sim$milliseconds as opposed to IR emission with timescales of $\sim$seconds \citep{Leger1988}. Recently, laboratory studies have reproduced RF as a cooling mechanism for PAH molecules \citep[e.g.,][]{Bernard2017, Bernard2023, Navarro_Navarrete2023, Stockett2023}, indicating that RF can be a powerful mechanism to significantly enhance the survival rate of small particles under high UV radiation densities. Another potential explanation for excess short wavelength PAH emission is the fact that larger PAHs can ``act'' like smaller PAHs in high radiation field environments. This arises due to the shorter mean time between photon absorptions in a high UV radiation density environment, leading PAH molecules to have higher vibrational energy, which in turn will emit at shorter wavelengths \citep{Draine2021}. The nature of the small grains and their energy-loss channels, along with their resilience under harsh conditions, will greatly impact the evolution of the ISM, the heating of gas via the photoelectric effect, and the UV/optical extinction law. JWST now allows us to probe significantly smaller physical scales than was previously possible, and trace the true variation in the 3.3 \um\ emission within galaxies over a wide range of environments.

\begin{figure}
    \centering
        \includegraphics[width=0.495\textwidth]{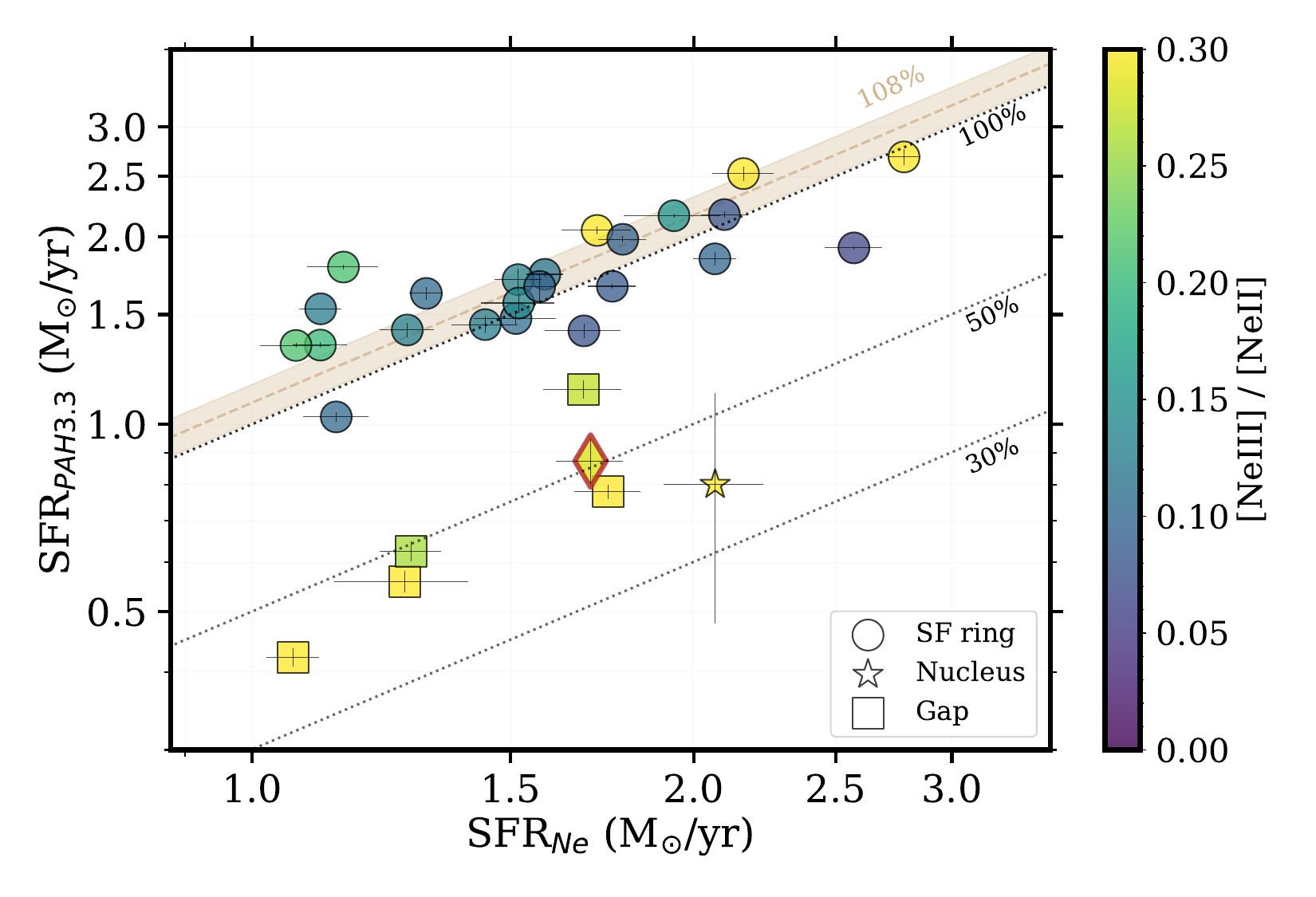} 
    \caption{The PAH 3.3-derived SFR vs. the neon-derived SFR in NGC~7469. The two SFR indicators follow each other within the uncertainties in star-forming cells. On average SFR$_\mathrm{PAH3.3}$ is slightly (8\%) higher compared to SFR$_\mathrm{Ne}$ as indicated by the 1:1 line (100\%). The shaded area indicates the 1-$\sigma$ range of the SF ring cells. However, the gap and the nucleus show significantly lower PAH-derived SFR, as expected, due to both weaker PAH emission and enhanced neon emission from the AGN. The red diamond represents the total extraction from the FoV of the IFU 
    (arbitrarily divided by 20 on both axes to allow it to be plotted with the starburst ring points) and shows a $\sim$50\% drop in 3.3\um\ PAH inferred SFR, as might be measured for NGC~7469 at high redshift and lower spatial resolution. 
    } 
    \label{fig:pah33_Ne_SFR}
\end{figure}

\subsection{PAH 3.3 as a SFR indicator}
JWST has opened up the possibility of spectroscopically detecting PAH emission in large numbers of high-z galaxies \citep{Spilker2023} and, in particular, using this feature as a star formation rate indicator. The 3.3\um\ PAH is the last band to shift beyond the long-wavelength coverage of MIRI, making it a prime tool to probe star formation out to redshifts of $z\sim6$. At low redshifts, it has been shown that there is a tight correlation between the SFR probed by PAH 3.3 and \NeII+\NeIII\ emission in metal-rich, star-forming galaxies; metal-poor galaxies, on the other hand, exhibit peculiar behavior that may lead to an underestimation of the SFR via the use of PAHs, by nearly an order of magnitude in some cases \citep{Lai2020}.

Figure~\ref{fig:pah33_Ne_SFR} shows the comparison between the SFR derived from the 3.3\um\ PAH \citep{Lai2020}, which can be written as

\begin{equation}
\mathrm{log} \bigg(\mathrm{\frac{SFR}{M_{\odot}yr^{-1}} \bigg) = -(6.80 \pm 0.18) + log \bigg(\frac{L_{PAH \ 3.3}}{L_{\odot}} \bigg)}
\label{eqn:SFR}
\end{equation}

\noindent and from the neon lines, which is based on Eq. 13 in \citet{Ho2007}, with the fraction of ionized neon emission derived from Eqn. (5) in \citet{Zhuang2019}. An independent test has also been conducted by comparing our PAH-derived SFR to the PAH~6.2 to IR luminosity relation reported in Table 4 of \citet{Gruppioni2016}. By assuming an average 6.2/3.3 PAH ratio of 2.87 found in our star-forming cells, the SFRs provided in \citet{Lai2020} and \citet{Gruppioni2016} are consistent at 10\% level.

In Figure~\ref{fig:pah33_Ne_SFR}, on average, the measurements in the SF ring follow the 1:1 line, showing slightly higher (8\%) SFR$_\mathrm{PAH3.3}$ than SFR$_\mathrm{Ne}$ with a small spread of 0.07 dex. However, the gap region cells and the nucleus show substantial deviations from the correlation due to the suppressed PAH emission in and around the AGN. An observation of a galaxy like NGC~7469 at high redshift, where the AGN and ring were spatially unresolved, would result in an underestimate of the star formation rate by about 50\% if measured from PAH 3.3 alone, due to the mixing of the AGN and starburst emission in the beam. 






\section{Summary}
With the advent of JWST, we can now probe the ISM in the vicinity of a powerful AGN on scales of a few hundred pc. In this \textit{Letter}, we present observations of the ISM in the nearby Seyfert galaxy NGC~7469 using NIRSpec IFU, supported by MIRI observations first reported in \citet{Lai2022}. In particular, we focus on the properties of small and neutral aromatic grains traced by the 3.3\um\ PAH band together with a nearby aliphatic feature at 3.4\um. Our findings can be summarized as follows:

\begin{itemize}
    \item With the NIRSpec IFU, we are able for the first time to confidently separate the dust features at 3\um\ in a resolved extragalactic source. In NGC~7469, the aliphatic (3.4\um) to aromatic (3.3\um) ratio varies by a factor of two across the IFU FoV, and this ratio is slightly lower near the AGN. This low 3.4/3.3 flux ratio, although with scatter, is most visible in the region between the Seyfert nucleus and the starburst ring where the \NeIII/\NeII\ line flux ratio reaches values of 0.5---1.0. The 3.4/3.3 ratio at the nucleus is very low compared to all other regions, suggesting extreme photo-destruction of the grains by the AGN.
    
    \item The 11.3/3.3 PAH ratio, which is a measure of the average grain size, is significantly larger ($>$2$\times$) in the vicinity of the AGN than it is in the starburst ring, suggesting a destruction of small grains within 300~pc of the nucleus, consistent with the results of \paperI. The nuclear spectrum displays a seemingly low 11.3/3.3 ratio, although with a large uncertainty as all the PAH fluxes are suppressed in this region.

    \item The total PAH-to-Ne ratio throughout the NIRSpec field is low in NGC~7469 compared to most local LIRGs and star-forming galaxies. However, the 3.3-to-Ne ratio and the fractional 3.3\um\ PAH power (3.3/$\Sigma$PAH) in the starburst ring of NGC~7469 is high compared to most LIRGs, which may be due to recurrent fluorescence of small grains or multiple photon absorption by large grains on 100~pc scales.
    
    \item An unresolved measurement of the PAH and Ne line emission in the central region of NGC~7469 that includes \emph{both} the AGN and the star-forming ring, such as might be observed with JWST at high redshift, would lead to an underestimation of its PAH-derived star formation rate by about 50\%. Thus, caution should be exercised when using PAHs to measure SFR in high-z AGN hosts or galaxies where AGN cannot be safely ruled out.
    
\end{itemize}

Here we have shown that the 3.3\um\ PAH feature varies on sub-kpc scales in NGC~7469, and that these variations can reveal changes in the grain populations as they react to changes in the ISM properties around an AGN and an intense starburst in this composite galaxy. Further observations of other nearby sources like NGC~7469 will be critical to fully understand, from a statistical standpoint, the expected variations in the PAH band fluxes and ratios that we can expect to see at high redshift with JWST in the coming years.


\begin{acknowledgments}
This work is based on observations made with the NASA/ESA/CSA \emph{JWST}. TSYL acknowledges funding support from NASA grant JWST-ERS-01328. The data were obtained from the Mikulski Archive for Space Telescopes at the Space Telescope Science Institute, which is operated by the Association of Universities for Research in Astronomy, Inc., under NASA contract NAS 5-03127 for JWST. 
Research at UCI by M.B. and V.U was supported by funding from program \#JWST-GO-01717, which was provided by NASA through a grant from the Space Telescope Science Institute, which is operated by the Association of Universities for Research in Astronomy, Inc., under NASA contract NAS 5-03127.
V.U further acknowledges partial funding support from NASA Astrophysics Data Analysis Program (ADAP) grants \#80NSSC20K0450 and \#80NSSC23K0750, and HST grants \#HST-AR-17063.005-A and \#HST-GO-17285.001.
The Flatiron Institute is supported by the Simons Foundation.
H.I. and T.B. acknowledge support from JSPS KAKENHI Grant Number JP21H01129 and the Ito Foundation for Promotion of Science.
AMM acknowledges support from the National Science Foundation under Grant No. 2009416.
ASE and SL acknowledge support from NASA grant HST-GO15472 and HST-GO16914. YS was funded in part by the NSF through the Grote Reer Fellowship Program administered by Associated Universities, Inc./National Radio Astronomy Observatory.
SA gratefully acknowledges support from an ERC Advanced Grant 789410, from the Swedish Research Council and from the Knut and Alice Wallenberg (KAW) Foundation.
FK acknowledges support from the Spanish program Unidad de Excelencia María de Maeztu CEX2020-001058-M, financed by MCIN/AEI/10.13039/501100011033.
KI acknowledges support by the Spanish MCIN under grant PID2019-105510GB-C33/AEI/10.13039/501100011033.
F.M-S. acknowledges support from NASA through ADAP award 80NSSC19K1096.
Finally, this research has made use of the NASA/IPAC Extragalactic Database (NED) which is operated by the Jet Propulsion Laboratory, California Institute of Technology, under contract with the National Aeronautics and Space Administration.

\vspace{5mm}
\facilities{\emph{JWST} (NIRSpec \& MIRI), MAST, NED}


\software{Astropy \citep{Astropy2013, Astropy2018},
          CAFE \citep{Marshall2007},
          \emph{JWST} Science Calibration Pipeline \citep{Bushouse2022},
          lmfit \citep{Newville2014},
          Matplotlib \citep{Hunter2007},
          Numpy \citep{VanderWalt2011},
          QFitsView \citep{Ott2012},
          SciPy \citep{Virtanen2020}
          }

\end{acknowledgments}



\bibliography{Lai_NGC7469}
\bibliographystyle{aasjournal}



\end{CJK*}
\end{document}